# A new method to quantify and reduce projection error in whole-solar-active-region parameters measured from vector magnetograms

David A. Falconer[1,2], Sanjiv K. Tiwari[1], Ronald L. Moore[1,2], Igor Khazanov[2,3]

1. NASA Marshall Space Flight Center, Huntsville, AL 35812, USA
2. Center for Space Plasma and Aeronomic Research, University of Alabama in Huntsville, Huntsville, AL 35899, USA
3. Archarithms Inc., Huntsville, AL 35801, USA
   **Corresponding author Email:** David.a.Falconer@nasa.gov

## Abstract:

Projection error limits the use of vector magnetograms of active regions (ARs) far from disk center. In this *Letter*, for ARs observed up to $60°$ from disk center, we demonstrate a method of measuring and reducing the projection error in the magnitude of any whole-AR parameter derived from a vector magnetogram that has been deprojected to disk center. The method assumes that the center-to-limb curve of the average of the parameter's absolute values measured from the disk passage of a large number of ARs and normalized to each AR's absolute value of the parameter at central meridian, gives the average fractional projection error at each radial distance from disk center. To demonstrate the method, we use a large set of large-flux ARs and apply the method to a whole-AR parameter that is among the simplest to measure: whole-AR magnetic flux. We measure 30,845 SDO/HMI vector magnetograms covering the disk passage of 272 large-flux ARs, each having whole-AR flux $>10^{22}$ Mx. We obtain the center-to-limb radial-distance run of the average projection error in measured whole-AR flux from a Chebyshev fit to the radial-distance plot of the 30,845 normalized measured values. The average projection error in the measured whole-AR flux of an AR at a given radial distance is removed by multiplying the measured flux by the correction factor given by the fit. The correction is important for both the study of evolution of ARs and for improving the accuracy of forecasting an AR's major flare/CME productivity.

# 1 Introduction

Full disk, high cadence observations of Helioseismic and Magnetic Imager (HMI: Schou et al. 2012) vector magnetograms allow us to consistently measure magnetic parameters of solar active regions (ARs) even when they are far from disk center. However, vector magnetograms away from disk center suffer from projection error. Projection error increases with distance from solar disk center. Based on a limited data sample of 36 AR vector magnetograms, Falconer et al. (2006) concluded that projection error in whole-AR magnetic flux measured from line-of-sight (LOS) magnetograms of ARs within $30°$ from solar disk center can be corrected by dividing it by $\cos^2\theta$, $\theta$ being heliocentric angle of the AR's position. Thus, for measuring AR magnetic flux, a LOS magnetogram can be used for ARs observed within $30°$ of disk center by applying the above correction, but additional projection error becomes significant beyond 30° and deprojection of vector magnetograms to disk center becomes necessary.

In Figure 1, we show HMI magnetograms, LOS and vertical-field components, for AR 11944 at seven different positions/times during its disk passage. The LOS component of AR magnetograms far from disk center shows fictitious neutral lines, which occur when there is a change of polarity in the LOS component of the magnetic field but not in the vertical component of the magnetic field. Fictitious polarities always occur on the limb side of sunspots far from disk center and are typically crescent-shaped as in Figure 1. These artifacts are a projection effect.

Several deprojection problems degrade the measurement of total magnetic flux. These include ambiguity resolution, large transverse field noise, and foreshortening.

The transverse magnetic field direction in a vector magnetogram has an 180° ambiguity. This comes from the transverse field direction measured from the linear polarization of light, which has only an 180° range. Several techniques have been developed to resolve this ambiguity (Metcalf et al 2006; Leka et al. 2009; Georgoulis 2012). We use the HMI definitive data, already disambiguated by the minimum energy method (Leka et al. 2009; Borrero et al 2011). We expect that disambiguation error increases with AR distance from disk center.

The noise in the LOS field in HMI vector magnetograms is 5-10 G while the transverse field noise is ~100 G (Liu et. al. 2012; Hoeksema et al. 2014). The farther the AR is from the disk center the greater the component of the observed transverse field in the vertical field of the deprojected magnetograms, increasing the noise in the (deprojected) vertical magnetic field. This produces an overestimate of the AR's total magnetic flux. This error increases with increasing radial distance of the AR. This transverse-field-noise effect can be clearly seen in Figure 1, where deprojected HMI active-region patches (HARPs) from original HARPs far from disk center show obvious light and dark gray noise in quiet regions, which does not occur when the original HARP is near disk center, as on 2014 January 8 in Figure 1.

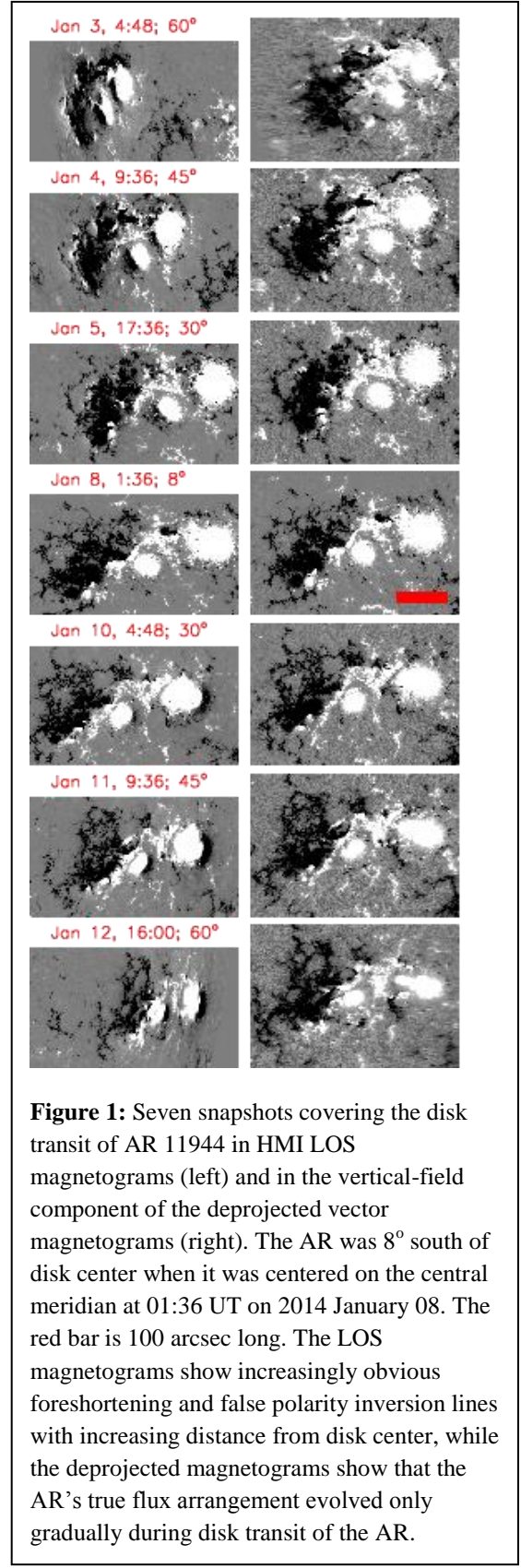

**Figure 1:** Seven snapshots covering the disk transit of AR 11944 in HMI LOS magnetograms (left) and in the vertical-field component of the deprojected vector magnetograms (right). The AR was 8° south of disk center when it was centered on the central meridian at 01:36 UT on 2014 January 08. The red bar is 100 arcsec long. The LOS magnetograms show increasingly obvious foreshortening and false polarity inversion lines with increasing distance from disk center, while the deprojected magnetograms show that the AR's true flux arrangement evolved only gradually during disk transit of the AR.

There is also projection error due to foreshortening, which causes an underestimate of the AR's total magnetic flux. This error results from flux of both polarities being in the pixels that have the polarity inversion lines in the LOS field. The farther AR from disk center, the larger the area of the Sun is covered by these pixels, and the more lost flux. By blurring these pixels, the point spread function (PSF) further increases this flux loss (Yeo et al 2014). The flux loss is further increased with distance from disk center by the increasing occurrence of false polarity inversion lines in the LOS magnetograms.

Some methods developed in the past to remove projection error by transforming the AR's vector magnetogram to disk center (Venkatakrishnan & Gary 1988; Venkatakrishnan et al. 1988; Gary & Hagyard 1990) retain some significant projection error for ARs beyond $45^o$ from disk center. An alternative way to cope with the projection error is to quantify the radial-distance dependence of the projection error in a measured whole-AR magnetic parameter and then correct for it. This is what we do in this *Letter,* i.e., we establish a method to quantify and reduce projection error in the magnitude (absolute value) of any whole-AR magnetic parameter measured from deprojected vector magnetograms. For demonstration of the method, we have used large-flux ARs (flux > $10^{22}$ Mx) and one of the simplest magnetic parameters, AR total magnetic flux. We have used HMI vector magnetograms in the present demonstration but the method should work for any vector magnetograms.

## 2  Measurement of Total Magnetic Flux

We use the MAG4 (short for Magnetogram forecast) tool to measure total magnetic flux of ARs. MAG4 is a code that automatically downloads and processes HMI magnetograms to forecast major flares, coronal mass ejections (CMEs), and solar proton events (Falconer et al 2011, 2012, 2014). In summer of 2015, MAG4 was upgraded to process vector magnetograms. Before that MAG4 used only LOS magnetograms. Now MAG4 first converts a vector magnetogram into vertical and horizontal field-vector components. This magnetogram is still in plane-of-sky pixels. MAG4 then resamples the magnetogram and creates square uniform pixels mapping the surface of the Sun at disk center using the method of Venkatakrishnan et al. (1988). Due to the relatively small size of HARPs, MAG4 does not take the spherical curvature of the Sun into account. We calculate the potential field from $B_z$ (Alissandrakis, 1981). In this *Letter,* we use a sample of 30,845 definitive-data HARP vector magnetograms. Although MAG4 measures many whole-AR magnetic parameters, here we establish the new method by concentrating on one parameter, AR total magnetic flux, measurement of which is the most straightforward and least erroneous. The total magnetic flux is chosen also because it is one of the slowest evolving parameters of a mid-life AR.

The total magnetic flux is given by

$$\Phi = \int |B_z|\, da.$$

where $B_z$ is the vertical magnetic field and the integral is over all pixels of the deprojected HARP with >100 G absolute field. The magnetic centroid of the deprojected HARP and the distance of that point from disk center in the plane-of-the-sky HARP are also determined. Since MAG4 analyzes only strong-field ARs, we use two other parameters measured from the deprojected HARPs to select strong-field ARs. One is magnetic area

$$A = \int da,$$

where the integral is over the same pixels as for $\Phi$, and the other is length of strong-field neutral line

$$L_S = \int dl,$$

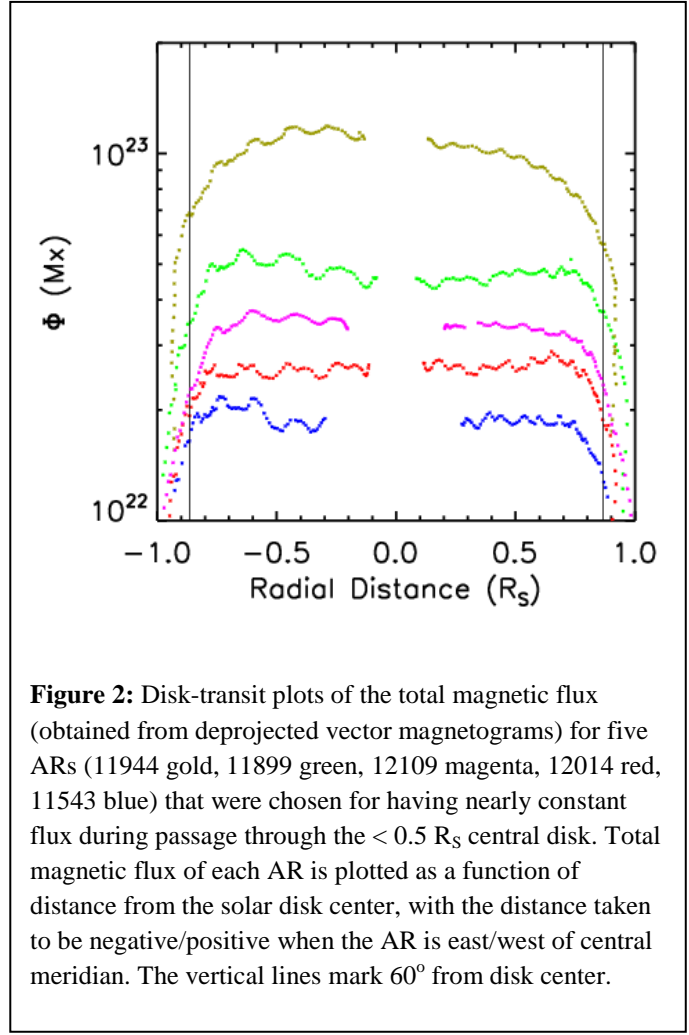

**Figure 2:** Disk-transit plots of the total magnetic flux (obtained from deprojected vector magnetograms) for five ARs (11944 gold, 11899 green, 12109 magenta, 12014 red, 11543 blue) that were chosen for having nearly constant flux during passage through the $< 0.5\ R_S$ central disk. Total magnetic flux of each AR is plotted as a function of distance from the solar disk center, with the distance taken to be negative/positive when the AR is east/west of central meridian. The vertical lines mark $60°$ from disk center.

where the integral is along all neutral-line intervals on which the potential horizontal magnetic field is greater than 150 G and separates opposite-polarity vertical magnetic field no weaker than 20 G.

For the present *Letter*, we use the MAG4 HMI database, which has only ARs that are observed by HMI and that qualify as so-called strong-field ARs. MAG4 uses only strong-field ARs in order for the horizontal field on much of the AR neutral line to be well above the noise level, for low-noise evaluation of free-energy proxies given by integrals of the horizontal field along the strong-field intervals of the neutral line. Following all of our previous work in developing MAG4 (Falcone et al 2011, 2012, 2014, and our foundational earlier work cited therein), we define a strong-field AR to be any AR for which the length $L_S$ of strong-field neutral line exceeds 75% of the square root of the area A of the vertical-field flux $\Phi$ of vertical field stronger than 100 G : $L_S/A^{0.5} > 0.75$.

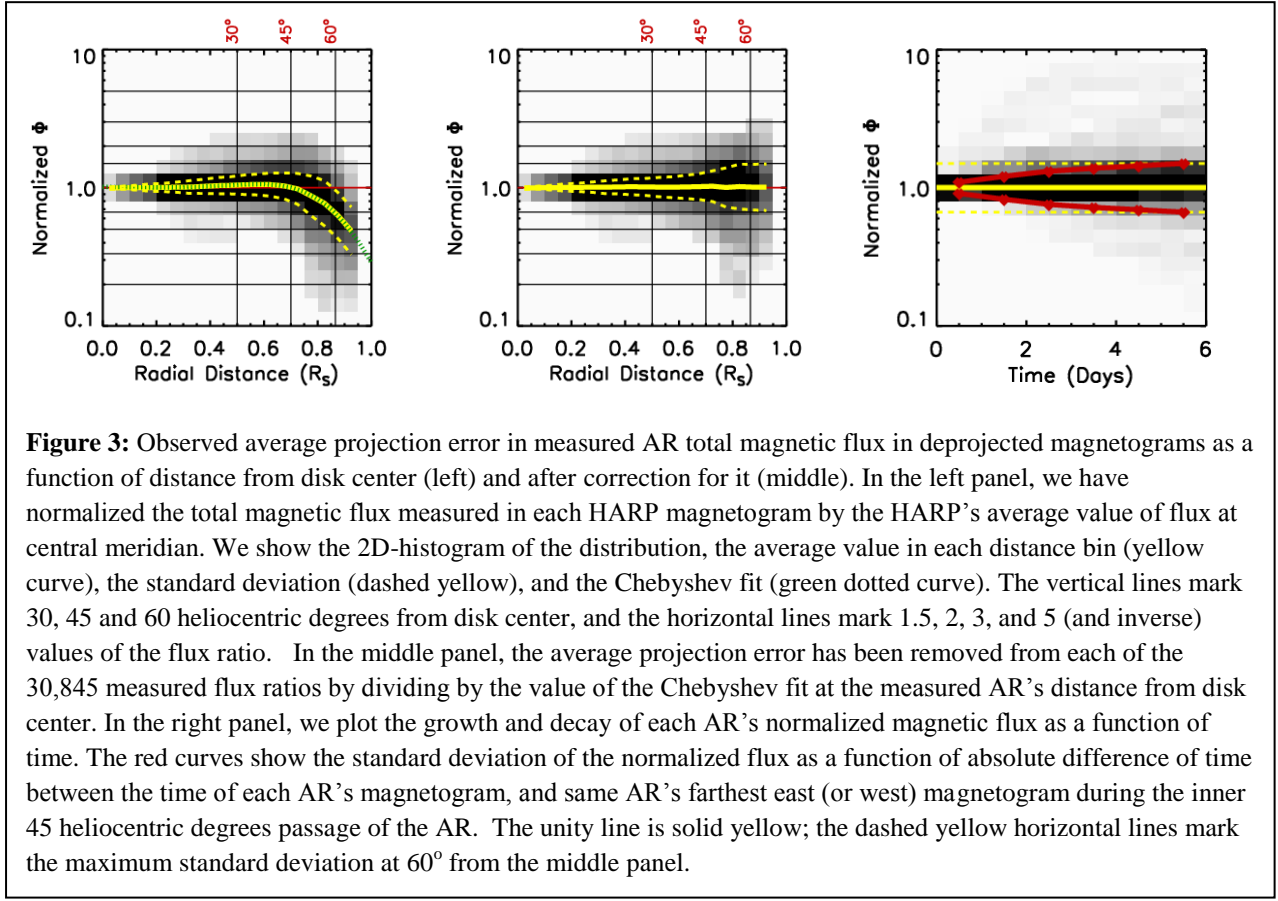

**Figure 3:** Observed average projection error in measured AR total magnetic flux in deprojected magnetograms as a function of distance from disk center (left) and after correction for it (middle). In the left panel, we have normalized the total magnetic flux measured in each HARP magnetogram by the HARP's average value of flux at central meridian. We show the 2D-histogram of the distribution, the average value in each distance bin (yellow curve), the standard deviation (dashed yellow), and the Chebyshev fit (green dotted curve). The vertical lines mark 30, 45 and 60 heliocentric degrees from disk center, and the horizontal lines mark 1.5, 2, 3, and 5 (and inverse) values of the flux ratio. In the middle panel, the average projection error has been removed from each of the 30,845 measured flux ratios by dividing by the value of the Chebyshev fit at the measured AR's distance from disk center. In the right panel, we plot the growth and decay of each AR's normalized magnetic flux as a function of time. The red curves show the standard deviation of the normalized flux as a function of absolute difference of time between the time of each AR's magnetogram, and same AR's farthest east (or west) magnetogram during the inner 45 heliocentric degrees passage of the AR. The unity line is solid yellow; the dashed yellow horizontal lines mark the maximum standard deviation at $60^o$ from the middle panel.

For this *Letter* only ARs with $\Phi > 10^{22}$ Mx are used. The $10^{22}$ Mx threshold was chosen because such large-flux ARs are the primary drivers of severe space weather. The same method can be applied to smaller-flux ARs, by using a large set of smaller-flux ($\Phi < 10^{22}$ Mx) ARs to obtain the center-to-limb correction curve for those ARs.

Note that we have measured whole HARPs for the analysis. Some of these HARPs contain more than one NOAA AR. We measure all of the strong-field regions in a HARP, treating them as a single AR. This does not affect the results presented in the current paper. Accordingly, in this paper we often refer to a HARP as an AR.

## 3  Radial dependence of Average Projection Error in AR Total Magnetic Flux and its Removal

The plots of measured magnetic flux $\Phi$ of five ARs through their disk transits are shown in Figure 2, each of which shows an upside-down U profile and an artificial 24-hour oscillation (Hoeksema et al 2014, and Couvidat et al 2016). Note that the X-axis is distance from disk center. The ARs in the east are given a negative distance so as not to have the east and west tracks overlie each other. Since none of these ARs transited through disk center, none of the ARs reached R = 0. For each AR, due to projection error from foreshortening, the measured values of the total magnetic flux are much smaller near the limbs than those near disk center. The selected five ARs showed only little evolution in total flux during central-disk passage. In other words, their flux did not rapidly grow or decay. The upside-down U profile is common to all ARs, including those that do show a real trend of growing or shrinking in size. The upside-down U profile makes using time derivatives of an AR's flux problematic for analysis. For example, for an AR far east of central meridian, we would not know if the AR's flux were actually growing as the AR rotated farther from the limb.

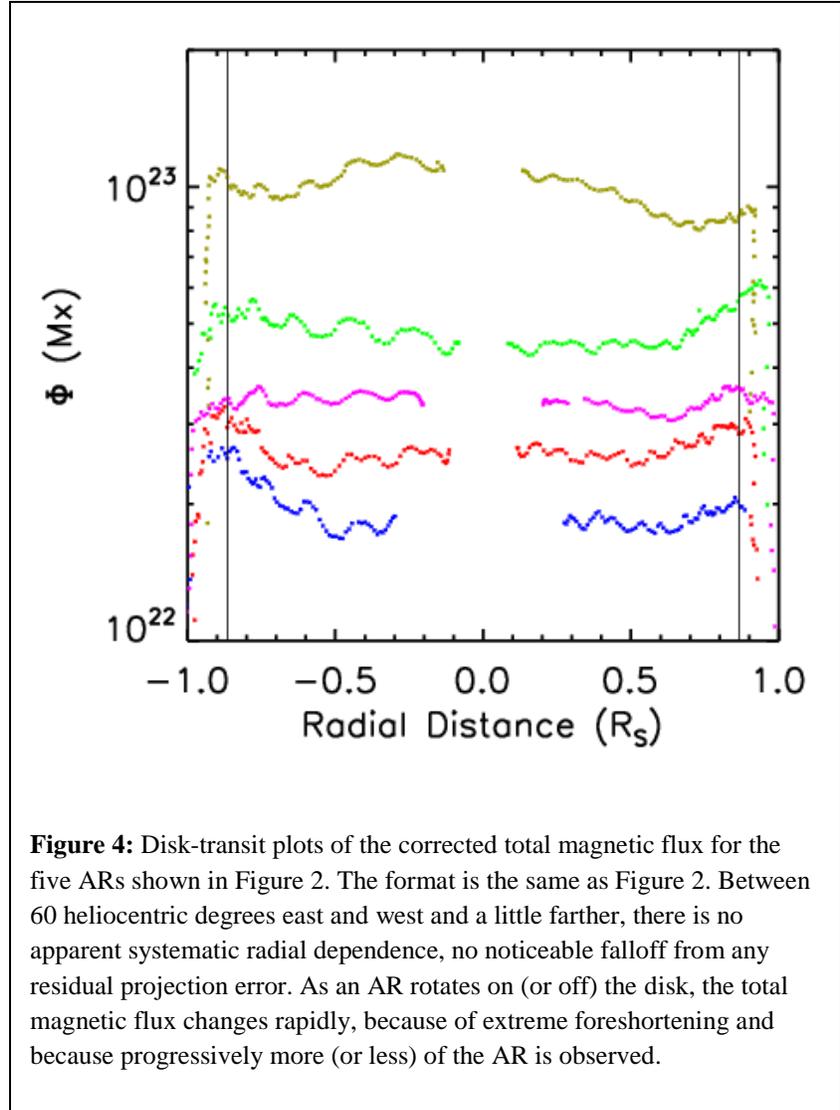

**Figure 4:** Disk-transit plots of the corrected total magnetic flux for the five ARs shown in Figure 2. The format is the same as Figure 2. Between 60 heliocentric degrees east and west and a little farther, there is no apparent systematic radial dependence, no noticeable falloff from any residual projection error. As an AR rotates on (or off) the disk, the total magnetic flux changes rapidly, because of extreme foreshortening and because progressively more (or less) of the AR is observed.

We quantify the average radial dependence of the projection error by using a large set of deprojected HARP magnetograms that yield disk-passage measurements of the total magnetic flux $\Phi$ of each of 272 HARPs. For each HARP, we average the measured flux in the strong-field area of that HARP over the HARP's magnetograms taken during the time when the magnetic centroid of the HARP's strong-field area is within 100" of central meridian. Then, each measured $\Phi$ from that HARP during its disk passage is divided by that average flux. Figure 3 (left and middle panels) shows a 2D histogram of the distribution of this flux ratio from a sample of 30,845 individual HARP magnetograms, all with strong-field flux of greater than $10^{22}$ Mx. Unlike in Figure 2, we leave the distance from disk center unsigned in Figure 3.

In Figure 3 (left panel), the yellow solid curve shows the average value of $\Phi$ in each of 20 equal radial-distance bins. It has a slight rise starting near 0.4 $R_S$ or 25$^o$ (before the 30$^o$ leftmost vertical line) and then falls back to 1 by 45$^o$ (the middle vertical line). At 60$^o$ the solid yellow curve shows that the uncorrected total magnetic flux is 2/3 of the true value. The dashed yellow lines show the 1σ standard deviation, (roughly 2/3 of the data points are within the two dashed lines). This standard deviation should not be thought of as error. As described later, nearly all of the spread is due to growth or decay of the ARs. Most ARs are not as steady as those shown in Figure 2.

In Figure 3 (left panel), a Chebyshev fit is done to all the data points, using the first five even terms (Press et al 1992). This fit is shown by the green dotted curve. The Chebyshev fit matches the bin-average curve and provides a continuous function, used for applying the correction. The fit shows that Φ is slightly overvalued in the ~ 30° – 45° distance from disk center, and is significantly undervalued beyond 50°. The values of the Chebyshev coefficients of the fit are 0.0049, 0.00087, -0.0016, 0.00084, and -0.000081.

Figure 3 (middle panel), shows the corrected distribution with the standard deviation of the corrected distribution around the average curve. The average of the corrected AR flux values shows hardly any deviation from 1.

The standard deviation of the corrected distribution grows with distance from disk center. The predominant cause of the growth of the standard deviation is AR evolution (Figure 3, right panel). To estimate the average growth/decay of the magnetic flux of our ARs, we use the same sample during the inner 45° disk passage where the correction factor is negligible. We normalized these not by central meridian passage, but by furthest east as well as furthest west measured AR Φ, so each HARP magnetogram between these two extreme positions is used twice. In effect we are treating the eastern most point as central meridian to simulate evolution in the western hemisphere, and treating the western most point as central meridian to simulate evolution in the eastern hemisphere. These are then folded, in a similar way as the other two panels of Figure 3. Absolute time relative to eastern most or western most magnetogram is plotted on the horizontal axis. This gives the flux evolution of our 272 ARs plotted in Figure 3, right panel. In 3-4 days, the standard deviation grows to the size seen at 60 heliocentric degrees in the Figure 3, middle panel. Comparison of the scatter in the right panel with that in the central panel of Figure 3 indicates that most of the scatter in the central panel is from AR evolution and that little (~20% or less) in the central panel is from variance in the projection error among the 272 ARs.

For the five AR HARPs of Figure 2, we show the corrected flux in Figure 4. Artificial 24-hour oscillation remains in the corrected flux. We can see that within 60° the ARs show no inverted-U trend during central disk passage. For tracking the evolution of an AR's total magnetic flux, this gives us confidence to use the magnetic flux measured from deprojected vector magnetograms of the AR out to 60° degrees, from which we have removed the radial-distance average projection error shown in Figure 3. Very close to the limb ($R > 0.95\ R_S$), all ARs show a rapid fall in Φ. This is due to extreme foreshortening that near the limb and to the ARs taking a finite time to rotate on or off the disk, effects that are not corrected by our method.

# 4 Discussion and Conclusions

A deprojected vector magnetogram of an AR is not exactly the same as the vector magnetogram of the AR if it were observed on the solar disk center. This is due to foreshortening (a pixel covers larger area towards limb), to the point spread function, to large transverse field noise that increases the noise in the vertical component of the deprojected magnetogram of ARs nearer the limb, and to error in $180^o$ ambiguity resolution for ARs near the limb. These combined effects result in measured AR uncorrected total magnetic flux being 2/3 of the true value at $60^o$ from disk center (Figure 3). Between $30^o$ - $45^o$ heliocentric angles there is a slight overvaluation of the total magnetic flux of an AR. Our interpretation for this is that transverse-field noise (and perhaps disambiguation error) causes the average curve in Figure 3 to rise with distance from disk center and that these projection effects are overcome and dominated by foreshortening error far enough from disk center, resulting in the fall of the average curve with increasing distance from disk center beyond about $40^o$. [This interpretation was confirmed by projecting a deprojected vector magnetogram of AR 11944 to different longitudes, converting back to LOS and transverse magnetic field maps, and applying the PSF to these maps, then deprojecting the magnetogram, and measuring the total magnetic flux. The center-to-limb curve obtained was qualitatively the same as that in the left panel of Figure 3 (showed the same trends) but was quantitatively different. To do this more correctly would require constructing the Stokes-parameter maps for the projected magnetograms with PSF applied (i.e., construct what the HMI vector magnetograph would have recorded).] The projection-error curve in Figure 3 was found by using a large sample of ARs, assuming that the average AR normalized true total magnetic flux is not a function of AR position on the disk. From this curve, the average projection error in an AR's total magnetic flux can be removed as in Figure 4. This correction is good out to at least 60 heliocentric degrees from disk center and is needed for improvement of scientific studies, for example, of evolutionary changes in AR magnetic flux.

Please remember that the correction curve we obtain from the method presented here is based on the average total flux of a large number of ARs during their disk-passage. The basic assumption is that the center-to-limb changes from the real evolution of ARs average out, e.g., that equal number of ARs grow or decay during their disk-passage. Given the large number of ARs in our sample, this assumption seems appropriate. To the degree that this assumption is true, an increase or decrease in the corrected $\Phi$ out to $60^o$ from disk center is due solely to the AR's growth or decay. We have also ignored the small east-west asymmetry in the HMI magnetograph (Hoeksema et al. 2014).

After the correction is applied, there is still the error due to the 24-hour artificial oscillation shown in Figures 2 and 4, which is a few percent and is practically negligible for some applications. This can be corrected by a fitting (e.g., see Hoeksema et al (2014) and Couvidat et al (2016)) that we do not apply here.

By correcting for the projection error and filtering out the 24 hour dependence, the magnitude of an AR's Φ can be well determined out to 60° from disk center. Most of the standard deviation in Figure 3, central panel, is due to AR evolution. The correction of projection-error shown in Figure 3 is especially needed if the AR's time derivative of Φ is of interest. Figure 3 indicates that without the correction applied only ARs within 45 heliocentric degrees can be used without much concern.

Although in this Letter we present the method by applying it to the simplest-to-measure magnetic parameter of an AR, the total magnetic flux, the method is suitable for correcting the magnitude (absolute value) of any whole-AR magnetic parameter, e.g., the various size, twist and free-energy proxies studied in Tiwari et al. (2015) and Bobra and Couvidat (2015). Correcting free-energy proxies out to 60° from disk center is needed for improving forecasting of AR's CME/flare productivity.

In conclusion, we have presented a new method to quantify and reduce the projection error in whole-active-region parameters that are measured from deprojected vector magnetograms. In this *Letter,* we have established the method by applying it to the total strong-field magnetic flux of 30,845 HARPs of 272 large-flux ($\Phi > 10^{22}$ Mx) ARs. We show that, with the correction applied, this parameter can be reliably used for scientific analysis of large-flux ARs observed up to 60° away from the solar disk center. We have found that the method works well for finding the center-to-limb correction curves for the absolute value of many other whole-AR magnetic parameters, including several free-energy proxies and signed parameters such as whole-AR magnetic twist. These curves will be presented in a more extensive future publication.

We thank the referee for leading us to make several improvements in the paper. Support for MAG4 development came from NASA's Game Changing Development Program, and Johnson Space Center's Space Radiation Analysis Group (SRAG). In particular, the authors want to gratefully acknowledge the continued support and guidance of Dan Fry (NASA-JSC) and David Moore (NASA-LaRC). DF acknowledges support from NextGen Federal Systems for the completion of this work. SKT is supported by an appointment to the NASA Postdoctoral Program at the NASA/MSFC, administered by USRA through a contract with NASA.